\documentclass[%
 reprint,
superscriptaddress,
 amsmath,amssymb,
aps,
]{revtex4-2}

\usepackage{graphicx}
\usepackage{dcolumn}
\usepackage{bm}
\usepackage{multirow}
\usepackage{makecell}
 \usepackage{float}

\usepackage{comment}
\usepackage{color}
\usepackage{url}

\usepackage[colorlinks]{hyperref}
\hypersetup{%
	plainpages=true,
	breaklinks=true,
	hypertexnames=false,
	pageanchor=true,
	colorlinks=true,
	linkcolor={blue},
	citecolor={red},
	urlcolor={blue},
	anchorcolor={black}
}

\usepackage{mleftright} 
\usepackage{styles_paper}

\begin{document}

\preprint{APS/123-QED}

\title{Universal control of a bosonic mode via drive-activated native cubic interactions}

\author{Axel M. Eriksson\orcid{0000-0002-2757-9804}}
\email{axel.eriksson@chalmers.se}
\affiliation{Department of Microtechnology and Nanoscience, Chalmers University of Technology, 412 96 Gothenburg, Sweden}

\author{Théo Sépulcre\orcid{0000-0002-2434-4487}}
\affiliation{Department of Microtechnology and Nanoscience, Chalmers University of Technology, 412 96 Gothenburg, Sweden}

\author{Mikael Kervinen\orcid{0000-0002-0164-9173}}
\affiliation{Department of Microtechnology and Nanoscience, Chalmers University of Technology, 412 96 Gothenburg, Sweden}

\author{Timo Hillmann\orcid{0000-0002-1476-0647}}
\affiliation{Department of Microtechnology and Nanoscience, Chalmers University of Technology, 412 96 Gothenburg, Sweden}

\author{Marina Kudra\orcid{0000-0003-4593-2859}}
\affiliation{Department of Microtechnology and Nanoscience, Chalmers University of Technology, 412 96 Gothenburg, Sweden}

\author{Simon Dupouy\orcid{0009-0002-8596-3887}}
\affiliation{Department of Microtechnology and Nanoscience, Chalmers University of Technology, 412 96 Gothenburg, Sweden}

\author{Yong Lu\orcid{0000-0002-7028-8316}}
\affiliation{Department of Microtechnology and Nanoscience, Chalmers University of Technology, 412 96 Gothenburg, Sweden}
\affiliation{Physikalisches Institut, University of Stuttgart, 70569 Stuttgart, Germany}

\author{Maryam Khanahmadi\orcid{0000-0002-3523-6449}}
\affiliation{Department of Microtechnology and Nanoscience, Chalmers University of Technology, 412 96 Gothenburg, Sweden}

\author{Jiaying Yang}
\affiliation{Department of Microtechnology and Nanoscience, Chalmers University of Technology, 412 96 Gothenburg, Sweden}

\author{Claudia Castillo-Moreno}
\affiliation{Department of Microtechnology and Nanoscience, Chalmers University of Technology, 412 96 Gothenburg, Sweden}

\author{Per Delsing\orcid{0000-0002-1222-3506}}
\affiliation{Department of Microtechnology and Nanoscience, Chalmers University of Technology, 412 96 Gothenburg, Sweden}

\author{Simone Gasparinetti\orcid{0000-0002-7238-693X}}
\email{simoneg@chalmers.se}
\affiliation{Department of Microtechnology and Nanoscience, Chalmers University of Technology, 412 96 Gothenburg, Sweden}

\date{\today}

\begin{abstract}
Linear bosonic modes offer a hardware-efficient alternative for quantum information processing but require access to some nonlinearity for universal control. 
The lack of nonlinearity in photonics has led to encoded measurement-based quantum computing, which relies on linear operations but requires access to resourceful ('nonlinear') quantum states, such as cubic phase states. In contrast, superconducting microwave circuits offer engineerable nonlinearities but suffer from static Kerr nonlinearity.
Here, we demonstrate universal control of a bosonic mode composed of a superconducting nonlinear asymmetric inductive element (SNAIL) resonator, enabled by native nonlinearities in the SNAIL element. We suppress static nonlinearities by operating the SNAIL in the vicinity of its Kerr-free point and dynamically activate nonlinearities up to third order by fast flux pulses. 
We experimentally realize a universal set of generalized squeezing operations, as well as the cubic phase gate, and exploit them to deterministically prepare a cubic phase state in 60 ns. Our results initiate the experimental field of polynomial continuous-variables quantum computing.

\end{abstract}

\maketitle

Quantum information processing relies on the capability to operate quantum superpositions between several quantum states in a large system. A mainstream approach to quantum computing is based on ensembles of coupled two-level systems~\cite{krinner2022, acharya2023a}. However, an alternative approach is to utilize bosonic modes~\cite{Gottesman2001, ma2021quantum, fluhmann2019}, also called continuous-variables (CV) modes. Each bosonic mode directly gives access to a large Hilbert space of quantum states, which could be used to redundantly encode and protect quantum information from errors in a hardware-efficient manner~\cite{Gottesman2001, ofek2016, hu_quantum_2019, lescanne2020, sivak_real-time_2023, reglade_quantum_2023}, or ,in principle, implement CV quantum computing~\cite{Lloyd1999}. However, to operate a CV mode, it is crucial to introduce a nonlinearity that allows one to universally manipulate the superpositions of the quantum states~\cite{ma2021quantum}. 

On the photonics platform, the main obstacle to quantum computing is the weakness of the accessible nonlinearities (relative to the intrinsic losses) in optical crystals. A common strategy in photonics is therefore to use readily accessible Gaussian operations such as squeezers and beamsplitters  in combination with non-Gaussian measurements, such as those performed by single photon detectors~\cite{bourassa_blueprint_2021, bartolucci_fusion-based_2023}. One approach to universal control is to utilize resourceful non-Gaussian input states~\cite{Bartlett2002, Mari2012,wu2020d} such as the cubic phase state to realize non-Gaussian gates by gate teleportation~\cite{Gottesman2001}. 

In contrast to the challenges in photonics, nonlinearities for CV modes in superconducting circuits are readily available, originating from Josephson junctions, that act as nonlinear, low-loss inductive elements~\cite{hofheinz2008}. These CV modes can be constituted by superconducting cavities weakly (dispersively) coupled to a qubit providing the universal controlability~\cite{hofheinz2009, heeres2015, Kudra2022,eickbusch2022}. For the superconducting modes the problem compared to the photonic domain is reversed, since the linear CV modes inherit a static, always-on Kerr nonlinearity from the hybridization with the Josephson junction~\cite{reagor2016}, which limits the fidelity of operations. Weakening the dispersive interaction reduces the inherited nonlinearity, but also leads to slower interactions which are typically on the order of a microsecond~\cite{Kudra2022,eickbusch2022}. Furthermore, despite considerable advances in superconducting qubits, the variety of noise channels introduced by the ancilla control qubit still limits the system performance~\cite{kervinen2023, sivak_real-time_2023}.

However, a strength of superconducting devices is the possibility of tailoring and \emph{in situ} tuning the nonlinearities with the external magnetic flux through superconducting loops. Especially, by arranging Josephson junctions into an asymmetric loop as in a Superconducting Nonlinear Asymmetric Inductive eLement (SNAIL)~\cite{frattini2017}, the Josephson potential becomes asymmetric. Hence, suitable tuning of the external flux enables the cancellation of the static Kerr nonlinearity while maintaining a strong third-order nonlinearity, making the SNAIL useful in quantum amplifiers~\cite{Sivak2019, fadaviroudsari2023}, reducing the static interaction in qubits~\cite{noguchi2020a} and linear couplers~\cite{chapman2023}.

\begin{figure*}[!t]
\includegraphics[width=\textwidth]{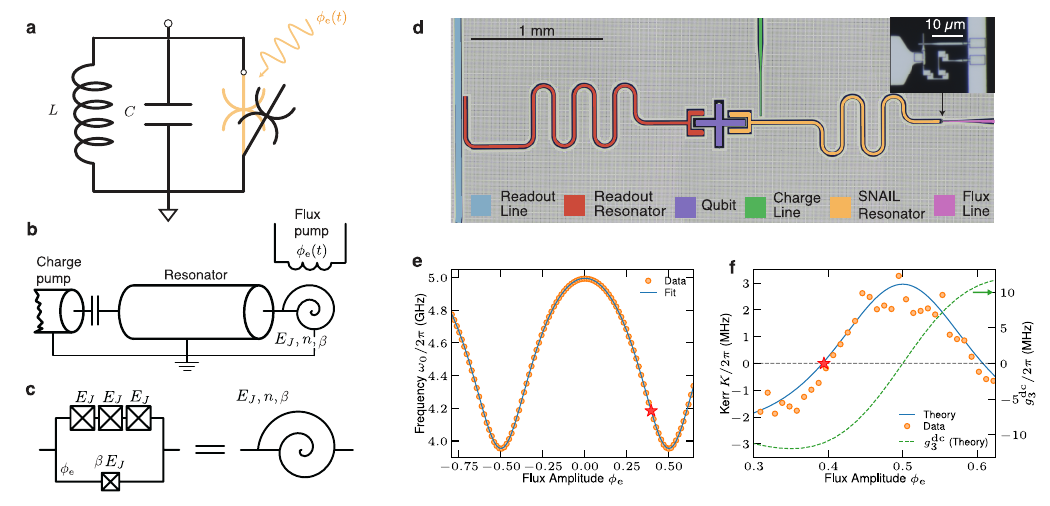}
\caption{\label{fig:setup}
Bosonic mode with drive-activated nonlinearities.
(a) Idealized circuit schematic. In the undriven state (open black symbol), the resonator behaves as a harmonic oscillator, characterized by its inductance $L$ and capacitance $C$. Manipulation of the quantum states is achieved by driving the resonator with flux $\phi_{\rm e}(t)$, which activates the (otherwise off-resonant) intrinsic nonlinearities. Different pulse compositions engage (closed orange symbol) a variety of interactions denoted by the switchable, purely nonlinear junction element. (b) Schematic of our circuit realization consisting of a SNAIL-terminated $\lambda/4$ coplanar waveguide resonator, following Ref.~\cite{Hillmann2020}.  (c) Schematic of the SNAIL element with $n=3$ Josephson junctions with Josephson energy $E_J$ and asymmetry factor $\beta$. (d) False-colored micrograph of the full device layout. The SNAIL-resonator is equipped with a charge and a flux drive line, and dispersively coupled to a transmon qubit with a readout resonator. (inset) Close-up of the SNAIL element with three large Josephson junctions on one arm, and a single junction on the other arm. (e) SNAIL-resonator frequency tuning vs static flux. The solid line is the fitted model taking into account the relevant microscopic parameters (see Methods). (f) Tuning of resonator nonlinearities vs static flux; fourth order (left axis) and third order (right axis). Effective Kerr $K^{(1)}$ and $g_3^{\rm ac}$ nonlinearities are predicted by the model fitted to the frequency tuning. The bosonic mode with drive-activated nonlinearities is realized and operated at the Kerr-free point marked with a star.
}
\end{figure*}

In this paper, we demonstrate universal control of a bosonic mode in a superconducting circuit comprising a planar resonator terminated by a SNAIL~\cite{Hillmann2020}, without an ancilla control qubit. The nonlinear controllability of the CV mode instead originates from intrinsic nonlinearities within the bosonic mode, which primarily manifest themselves when driven. Hence, different nonlinear interactions are turned on only when certain drive pulses are applied (Fig.~\ref{fig:setup}a). These nonlinearities are realized by embedding a SNAIL into the bosonic mode resonator itself (Fig.~\ref{fig:setup}b-c). By flux and charge driving the SNAIL-terminated resonator, we activate Gaussian and non-Gaussian (nonlinear) interactions to implement fast universal quantum gates within tens of nanoseconds. In contrast to previously demonstrated bosonic gate sets~\cite{heeres2015,eickbusch2022}, the gate set implemented here is a natively polynomial set including the non-Gaussian cubic gate. We thereby implement a universal gate set in the notion of CV quantum computing inspired by the original proposal by Lloyd and Braunstein~\cite{Lloyd1999}. Furthermore, by leveraging the native cubic gate, we experimentally generate a cubic phase state with enhanced speed and fidelity compared to our previous work~\cite{Kudra2022} in 3D cavities.

\bigskip
We implement our device on a superconducting planar architecture fabricated with conventional lithography techniques and measured at $\sim10$mK in a dilution refrigerator (see Methods). We realize the bosonic mode with drive-activated nonlinearities by terminating a $\lambda/4$ resonator with a SNAIL element at its current anti-node (micrograph in  Fig.~\ref{fig:setup}d). A transmon qubit \cite{koch2007} is dispersively coupled to the bosonic mode as well as to a readout resonator to perform Wigner tomography of the bosonic state~\cite{heeres2015}. Crucially, the qubit does not participate in the control of the bosonic mode. Importantly, the SNAIL element alters the boundary condition of the otherwise linear mode and thereby introduces the nonlinearities which will be utilized for the control. The flux tunability of the SNAIL potential results in a nonlinear frequency tuning of the mode with respect to the external magnetic field through the SNAIL flux loop, as measured via the cavity Ramsey protocol~\cite{Reinhold2019} (Fig. \ref{fig:setup}e). 

The nonlinear Hamiltonian of the SNAIL-terminated resonator~\cite{Hillmann2020}, driven by charge and flux, can be expanded around the minimum of the potential well and takes the form 
\begin{equation} \label{eq:hamiltonian}
\begin{split}
	\hat{H}/\hbar
	&= \omega_0 \hat{a}^\dagger\hat{a}
	+ \xi(t) \left(\hat a ^\dagger + \hat a \right) \\ 
  &+\sum_{n=1}^{\infty} \Big(g_n^{\rm dc}(\phi_{\rm e}^{\rm dc}) + g_n^{\rm ac}(\phi_{\rm e}^{\rm dc}) \phi_{\rm e}^{\rm ac} f(t)\Big) \left(\hat{a}^\dagger + \hat{a}\right)^n 
\end{split}
\end{equation}
where $g_1^{\rm dc} = g_2^{\rm dc} = 0$ 
and with frequency $\omega_0$, bosonic creation operator $\hat a^\dagger$, charge drive amplitude $\xi(t)$, and linear/nonlinear coefficients $g_i^j$ for $i\in \mathbb{N}$ and $j$ spanning both static dc terms as well as flux ac components, which depend on the static flux $\phi_{\rm e}^{\rm dc}=\Phi_{\rm e}^{\rm dc}/\Phi_0$, $\Phi_0=h/2e$. The flux drive field is constituted by the flux amplitude $\phi_{\rm e}^{\rm ac}$ and normalized pulse shape $f(t)$, where ${\rm max}(f(t))=1$. By fitting the microscopic parameters of the static components (see Methods) to the measured nonlinear tuning of the frequency as a function of static magnetic flux (Fig. \ref{fig:setup}e), we predict the strengths of the nonlinearities $g_i^j$ (Fig. \ref{fig:setup}f). The predicted effective Kerr is validated by measuring the same via the out-and-in protocol (see Methods). All fitted and characterized parameters are summarized in a table in the Supplementary materials.

Importantly, the magnitudes of the effective static Kerr nonlinearities are strongly suppressed at flux $\phi_{\rm e}^{\rm dc}=0.3930$, creating a Kerr-free region for phase space coordinates $|\alpha|\lesssim1.5$ (see Methods).  This cancellation is a key feature of the SNAIL, which is possible due to its asymmetric design allowing for the cancellation of one arbitrary order in the Taylor expansion of the potential~\cite{frattini2017}. Therefore, we choose this (close to) Kerr-free point as the operating point of our device, characterized by a resonant frequency $\omega_0/2\pi=4.158$ GHz, energy relaxation time $T_1=\SI{28}{\micro\second}$, and cubic nonlinearity $g_3^{\rm ac}/2\pi\sim-10$ MHz. Hence, when no drives are applied, the resonator behaves approximately as a harmonic oscillator. We achieve quantum control of the cavity state by sending microwave pulses to on-chip flux and charge lines. By judiciously choosing the frequencies and composition of the driving pulses, we activate a range of linear and nonlinear interactions, as demonstrated below.  

\begin{figure*}[!t]
\includegraphics[width=1\textwidth]{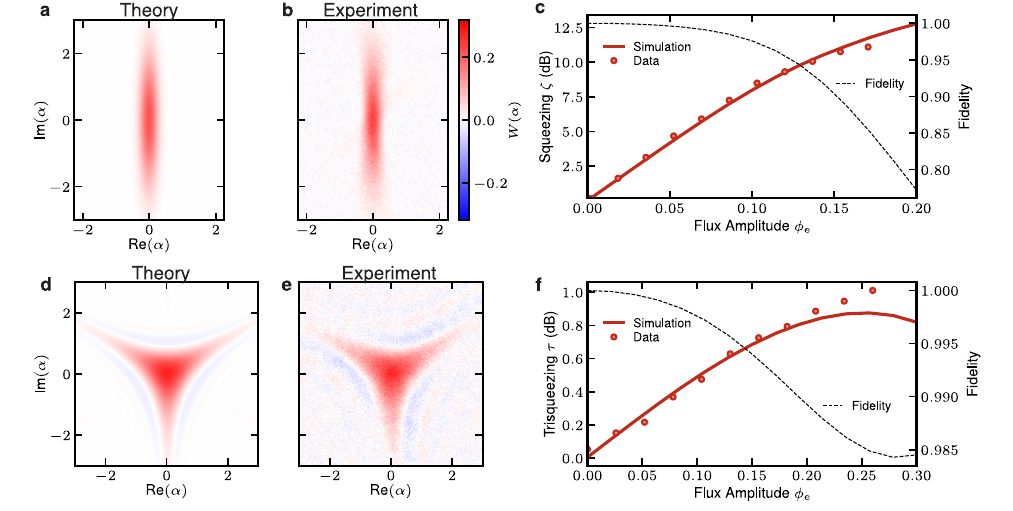}
\caption{\label{fig:interactions} Universal control of the SNAIL-resonator achieved by monochromatic flux pulses. Results are shown for squeezing (top panels) and trisqueezing (bottom panels). A 20 ns $2\omega_0$ pulse parametrically activates the squeezing interaction and squeezes the cavity vacuum state to $\zeta=-0.99$ (-8.63 dB). The resulting Wigner state is fitted with a pure squeezed state (a) via the Wigner overlap to the measured state (b). 
(c) Increasing squeezing level with flux pulse amplitude. Pulse duration also increases the squeezing level but is here fixed to 40 ns. Data points are obtained by fitting the measured Wigner functions to the corresponding pure states via Wigner overlap. Simulated state squeezing level (solid) and fidelity (dashed). Similarly, a 60 ns $3\omega_0$ flux pulse activates the trisqueezing interaction which trisqueezes the cavity vacuum state into a Wigner negative trisqueezed state fitted (d) and measured (e) with $\tau=-0.13$. (f) Increasing trisqueezing level analogous to the squeezing case.  
}
\end{figure*}

\bigskip
\label{univ}
The notion of universality introduced by Lloyd and Braunstein for continuous-variable systems \cite{Lloyd1999} requires a set of Gaussian gates and at least one non-Gaussian gate, which can be chosen arbitrarily among the polynomials of degree three or higher in the quadratures of the bosonic modes. Hence, we experimentally obtain a universal gate set by flux driving the SNAIL in such a way that each gate consists of a single monochromatic pulse at a multiple of the resonator frequency. We demonstrate the gate set by flux pumping the SNAIL at one, two, and three times the resonance frequency $\omega_0$, which resonantly engage the displacement, squeezing (Fig. \ref{fig:interactions}a-c) and trisqueezing~\cite{Chang2020} (Fig. \ref{fig:interactions}d-f) interactions, respectively. Together with the trivial rotation, these interactions constitute the universal generalized-squeezing gate set
\begin{align} \label{eq:disp}
\hat R(\theta) =\ & e^{-i\theta\hat a^\dagger \hat a}\\\label{eq:rot}
\hat D(\alpha) =\ & e^{\alpha \hat a^\dagger - \alpha^* \hat a}\\\label{eq:sq}
\hat S(\zeta) =\ & e^{(\zeta \hat a^\dagger{}^2 - \zeta^* \hat a^2)/2}\\\label{eq:trisq}
\hat T_s(\tau) =\ & e^{\tau \hat a^\dagger{}^3 - \tau^* \hat a^3}.
\end{align}
with rotation $\hat R(\theta)$ by an angle $\theta$,  displacement $\hat D(\alpha)$, squeezing $S(\zeta)$ and trisqueezing $T_s(\tau)$, where the complex parameters $\alpha$, $\zeta$, and $\tau$ describe the magnitude and phase of the operations. The magnitude of $\alpha$, $\zeta$, and $\tau$ are to first order proportional to the Hamiltonian parameters $g_1^{\rm ac},\ g_2^{\rm ac}$ and $g_3^{\rm ac}$ in equation (\ref{eq:hamiltonian}), respectively (see Methods). Crucially, the trisqueezing interaction is non-Gaussian, promoting the otherwise Gaussian gate set to a universal gate set. By engaging the trisqueezing gate, we produce the non-gaussian Wigner-negative~\cite{Mari2012} trisqueezed state (Fig.~\ref{fig:interactions}e). The quantum states are measured by direct Wigner tomography via the dispersively coupled spectator qubit. The squeezing and trisqueezing levels can be enhanced until $\sim$9dB and $\sim$0.8dB, respectively, above which the simulated fidelities start to drop because of higher-order nonlinearities. Furthermore, these flux-activated gates are executed within tens of nanoseconds. This is to be contrasted with bosonic gates assisted by a dispersively coupled qubit, such as selective number-dependent arbitrary phase (SNAP) gates \cite{heeres2015,Kudra2022} or optimal control pulses \cite{heeres2017}, which are 10 to 100 times slower due to the weakness of the dispersive interaction. 

We emphasize that the gates demonstrated in Fig.~\ref{fig:interactions} are performed by driving flux tones, whereas experimental demonstrations for superconducting 3D modes often are limited to charge driving because of the difficulty to bring ac-flux fields into the superconducting cavities~\cite{lu2023}. This enables us to use interactions via both the linear charge and parametric flux paradigm to tailor a variety of gates (see Methods for illustrative diagrams). 
As an example, a trisqueezing gate could in principle be performed solely with charge driving at $\omega_{\rm d} = 3\omega_0$ in presence of a $g_4^{\rm dc}$ interaction term. But this interaction also triggers a static Kerr effect.
Tuning the system to reach the Kerr-free-flux point is not beneficial in this context, as it also cancels the amplitude of trisqueezing gate.
By contrast, the use of a parametric drive $g_3^{\rm ac}$ does provide the trisqueezing gate at the Kerr free point, without any residual interaction when idling. Similarly, each $g_n^{ac}$ in equation~(\ref{eq:hamiltonian}) allows one to activate an $n$-photon process by flux driving resonantly at $n\omega_0$, with greatly suppressed unwanted static effects. 

\begin{figure*}[!t]
\includegraphics[width=1\textwidth]{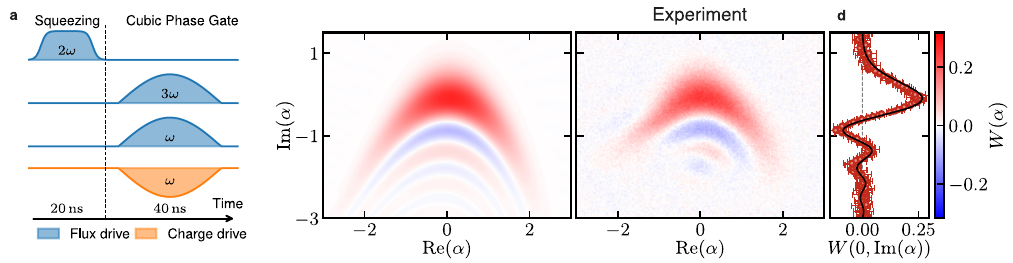}
\caption{\label{fig:cubic} Experimental demonstration of the cubic gate by generation of a cubic phase state in 60 ns. (a) Pulse sequence. 
First, we apply a squeezing gate to the vacuum by a \SI{20}{\nano\second} $2\omega_0$ flux pulse and then a cubic gate constituted by a $3\omega_0$ flux pulse, a $1\omega_0$ flux pulse and a counteracting $1\omega_0$ charge pulse, at the same time, for \SI{40}{\nano\second}. (b) Fitted and (c) measured Wigner function of the generated cubic phase state 
with squeezing $\zeta=-0.61\ (5.3~{\rm dB})$ and cubicity $\gamma=0.11$ and a residual displacement $\alpha=-0.090+i0.025$. (d) Line cut of the Wigner function in (c) at ${\rm Re}(\alpha)=0$: experimental data (symbols), and corresponding theory prediction for the state in (b) (solid line). 
}
\end{figure*}

\bigskip
Following the proposal by Hillmann et al.~\cite{Hillmann2020}, we utilize simultaneous flux and charge drives to natively implement the cubic phase gate (Fig.~\ref{fig:cubic}), 
\begin{equation}
     \hat C(\gamma) = e^{i \gamma\left(\frac{\hat a^\dagger{} + \hat a}{\sqrt{2}}\right)^3}
\end{equation}
with cubicity parameter $\gamma$. 
The cubic interaction contains the cross terms proportional to $\hat a^\dagger{} \hat a^2$ and $\hat a^\dagger{}^2 \hat a$, which distinguishes it from the trisqueezing gate. To activate the full cubic interaction (Fig.~\ref{fig:cubic}a), we therefore simultaneously apply a $3\omega_0$-flux pulse which captures the trisqueezing interaction and a $1\omega_0$-flux pulse capturing the cross terms. However, since the $1\omega_0$-flux pulse also engages direct displacement, we concurrently apply a $1\omega_0$-charge pulse to cancel the linear displacement. Since the displacement interaction is much stronger than the cross interaction ($g_1^{\rm ac}/g_3^{\rm ac}\approx2100$), the cubic phase gate requires precise experimental calibrations to balance the timing, strength and angles of the drive pulses (see Methods).

Finally, we combine the squeezing gate with the cubic gate to deterministically generate a
cubic phase state from the vacuum state.
We generate the resourceful~\cite{Konno2021} cubic phase state $|\zeta, \gamma \rangle = \hat C(\gamma)\hat S(\zeta)|0 \rangle$ from the vacuum state $|0 \rangle$ by first applying a 20 ns squeezing gate followed by a 40 ns cubic gate. This sequence results in a cubic phase state (Fig.\ref{fig:cubic}b-c) characterized by $\zeta=-0.61$ (corresponding to 5.3 dB of squeezing) and $\gamma=0.11$. The fidelity to the closest pure cubic phase state is estimated to be 92 \% from generative adversarial neural network reconstruction~\cite{ahmed2021} (see Methods).
Importantly, the cubicity of the cubic phase state showcased in this study can be continuously increased by simply employing a longer cubic gate. Being able to enhance the cubicity via the cubic gate is in striking contrast to the prior demonstration of a cubic phase state~\cite{Kudra2022}, in which selective-number arbitrary phase (SNAP) and displacement gates had to be re-optimized for every new cubicity value.

We construct an error analysis  using numerical simulations to examine the factors that restrict the performance of cubic phase state generation (see Methods).
The analysis reveals that the fidelity of the generated state is primary limited by the coherence time of the SNAIL resonator ($T_2^*=\SI{2.8}{\micro\second}$ dominated by 1/f noise, see Methods), followed by residual thermal population in the SNAIL-resonator ($n_{\rm th}$ = 2.4\%). By contrast, simulations indicate that the higher order corrections $g_5^{\rm dc}$ and $g_6^{\rm dc}$ have a very limited influence on the fidelity of the produced cubic phase state, whereas they become a dominant source of error for larger states.
In future realizations, the flux sensitivity of the SNAIL at the operational Kerr-free point could potentially be improved by optimizing the SNAIL parameters~\cite{chapman2023} without compromising the strengths of the desired nonlinearities. Thermal population could be further lowered by improving the filtering of the flux line (see Supplementary materials). A single SNAIL could be replaced by an array of $m$ SNAILs, for which higher-order nonlinearities at order $n$ are suppressed as $g_n(t) \sim {}m^{1-n}$. Care must be taken to ensure the homogeneity of both the SNAILs within the array and the flux applied to each SNAIL. Simulations also indicate that there is room to drive the device harder before other unwanted nonlinear effects occur. Faster gates could be obtained by removing the hardware limitations that prevented us from further increasing the amplitude of the driving tones in the present experiment. Finally, we observe a systematic flux drift in the data batches of the cubic phase state. The drift manifests itself as a tilting cubic phase state, which we ascribe to the system drifting away from the Kerr-free point. Hence, we expect more frequent system calibration to improve the state generation performance.

\bigskip
In conclusion, we have experimentally demonstrated a polynomial gate set for a bosonic mode in the notion proposed by Lloyd and Braunstein~\cite{Lloyd1999}. Hence, our platform initiates the field of continuous-variable–noisy intermediate scale quantum (CV-NISQ) polynomial quantum computing~\cite{Arrazola2019} and simulations~\cite{Anderson2022}. We envision utilizing the SNAIL-resonator as a specialized computational unit interfacing with longer-lived bosonic modes (for example, via beamsplitter interactions \cite{chapman2023}), either to rapidly prepare highly nonclassical states, or to execute highly nonlinear gates. In addition, our demonstration of a strong, native trisqueezing gate spurs the question how continuous-variable algorithms could directly benefit from such a resource. 
The tunable nonlinearities in the SNAIL-resonator further allow for switching its functionality \textit{in situ} to operate in the Kerr-cat regime~\cite{puri2017, xu2022, iyama2023} and even investigate higher order time crystals~\cite{Wilczek2012, Zhang2017}. 
Another direction would be to combine the intrinsic polynomial interactions with the control offered by an ancilla qubit via optimized selective number-dependent arbitrary phase gates~\cite{Kudra2022} or full optimal control~\cite{heeres2015}.
Finally, if the fast progress on efficient microwave-to-optics transducers~\cite{tu_high2022} continues, one can envision upconverting cubic phase states produced on the microwave platform to optical frequencies. That is, generating the resource states required for fault-tolerant measurement-based quantum computing on a hybrid superconducting-photonics platform. 

\section{\label{sec:methods}Methods}

\subsection{Estimating microscopic parameters}
The Hamiltonian used to represent the SNAIL resonator in equation~(\ref{eq:hamiltonian}) is written in terms of parameters $\omega_0$, $g^{\rm ac}_i$ and $g^{\rm dc}_i$. It would be difficult to extract a value from each of them independently. 
Instead, we express them as functions of a few relevant microscopic parameters of the underlying circuit, which are obtained all at once by fitting the $\omega_0 = f(\phi_{\rm e})$ experimental data. We now describe these microscopic parameters.

The SNAIL element is described in Fig.~\ref{fig:setup}.d. 
We denote $\varphi$ the superconducting phase at one of its node, the other being grounded.
If we assume our probing frequency negligible compared to the plasma frequencies of the SNAIL junction array, we can model it as a nonlinear potential imposed on $\varphi$:
\begin{equation}
	U(\varphi) =
	- E_{\rm J} \left(\beta \cos(\varphi)
	+ n \cos\left(\frac{\phi_{\rm e} - \varphi}{n}\right) \right),
\end{equation}
where $\phi_{\rm e}$ is the external flux piercing the SNAIL loop (From now on, $\hbar = 2e = 1$).
The potential minimum $\varphi_{\rm m}$ is given by $\beta\sin(\varphi_{\rm m})=\sin((\phi_{\rm e} - \varphi_{\rm m})/n)$.
Expanding $U$ around this minimum as a Taylor expansion, $U = \sum_{n=2}^\infty \varphi^n \mathrm{d}^n U/\mathrm{d}\varphi^n(\varphi_{\rm m})/n!$.
All terms are non-vanishing, contrary to what would happen for a SQUID element. 

The bare resonator (without the SNAIL) is characterised by its lowest mode frequency $\omega_{\rm \infty}$ and impedance $Z$. 
The quadratic term in the Taylor expansion will act like a boundary inductor, shifting the resonance frequency to $\omega_0$. 
The SNAIL flux is expanded in the resonator creation/annihilation operators as $\hat\varphi = \Phi (\hat{a}^\dagger + \hat{a})$, with proportionality factor $\Phi$, such that the resonator Hamiltonian reads
\begin{equation}\label{eq:hamiltonian_derivatives}
	\hat{H} = \omega_0 \hat a^\dag \hat a
	+ \sum_{n=3}^\infty \frac{\Phi^n}{n!} \derivat{n}{U}{\varphi}{\displaystyle \varphi_{\rm m}} (\hat a^\dag + \hat a)^n.
\end{equation}
See Supplementary Materials for expressions of $\omega_0$ and $\Phi$ as functions of the microscopic parameters. 
$\omega_0$ is measured \textit{via} a Ramsey interference protocol.
It is fitted against our microscopic model with $5$ free parameters, $\beta$, $\omega_\infty$, $Z/E_{\rm J}$, and external flux calibration with a linear model, $\phi_{\rm e}^{\rm dc} = V^{\rm dc}/V^{\rm dc}_{0}+ \phi_{\rm offset}^{\rm dc}$, with voltage $V^{\rm dc}$ applied to the flux source, voltage corresponding to one flux quanta $V^{\rm dc}_{0}$ and the flux offset $\phi_{\rm offset}^{\rm dc}$.
The fitted curve is represented in Fig.~\ref{fig:setup}e, and the fit parameter values are given in Supplementary Materials. $E_{\rm J}$ is obtained independently by normal state resistance measurement.

\subsection{AC and DC nonlinear terms} The external flux is modulated around its DC value as $\phi_{\rm e}(t) = \phi_{\rm e}^{\rm dc} + \phi_{\rm e}^{\rm ac} f(t)\cos(\omega_{\rm d}t)$, where $f(t)$ is a slowly varying pulse envelope with ${\rm max}f(t)=1$, which drives the system. 
Accordingly, we separate the coefficients in equation~(\ref{eq:hamiltonian_derivatives}) in their DC and AC parts.
We expand the Hamiltonian coefficients at first order around $\phi_{\rm e}^{\rm dc}$ (assuming $\phi_{\rm e}^{\rm ac} \ll 1$) and obtain equation~(\ref{eq:hamiltonian}),
where $g_1^{\rm dc} = g_2^{\rm dc} = 0$ and
\begin{equation}
	g_n^{\rm dc}
	= E_{\rm J} \frac{\Phi^n}{n!} \left.\frac{\partial^n U}{\partial \varphi^n} \right\rvert_{\overset{\displaystyle \phi_{\rm e}^{\rm dc}}{\displaystyle\varphi_{\rm m}}},
	\quad g_n^{\rm ac}
	= E_{\rm J} \frac{\Phi^n}{n!} \left. \frac{\partial}{\partial \phi_{\rm e}}\frac{\partial^n U}{\partial \varphi^n} \right\rvert_{\overset{\displaystyle \phi_{\rm e}^{\rm dc}}{\displaystyle\varphi_{\rm m}}}.
\end{equation}
Expressions for the derivatives can be evaluated explicitly.
Note that nonlinear coefficients $g_n^{\rm dc}$ and $g_n^{\rm ac}$ are completely determined by previously fitted parameters $\omega_{\infty}$, $\alpha$, $Z$ and $E_{\rm J}$, as done in Fig.~\ref{fig:setup}.f.

\subsection{Numerical simulations} \label{sec:numerical_simulations}
We systematically compare the performances of our experimental system with numerical simulations of its dynamics.
The simulations are done with the \texttt{QuTiP}~\cite{johansson_2012_QuTiPOpensource} \texttt{Python} library, using time integration of the Lindblad equation of motion. 
We include all drives with their pulse shapes as, nonlinear terms up to order $6$ in equation (\ref{eq:hamiltonian}), energy relaxation and dephasing, all according with the numerical values obtained from experimental data summarized in Supplementary materials.
We assume the initial cavity state and the environment to be in thermal equilibrium at $\SI{50}{\milli\kelvin}$, a value obtained by measuring residual excited population of the resonator.
These simulations are used in Fig.~\ref{fig:interactions}c and f, where they provide a theory to experiment agreement beyond the linear regime at weak drive.

\subsection{Kerr-free point} The rotating wave approximation (RWA) is a customary method to study driven nonlinear oscillators~\cite{rasmussen_2021_SuperconductingCircuit}, which neglects fast rotating terms in the oscillator frame. 
The method can be systematically extended in a perturbative expansion in $g_n/\omega_0$~\cite{james_2007_EffectiveHamiltonian,Hillmann2022,venkatraman_2022_StaticEffectiveb}. 
\begin{table}
	
	\caption{Perturbative expansions of static and driven nonlinear processes. 
    Each term is represented by a representative diagram, in spirit of~\cite{frattini_2018_OptimizingNonlinearity,Hillmann2022,xiao_2023_DiagrammaticMethod}, which aids exhausting all contributions. 
    A vertex with n straight legs represents $g_n$. 
    An extra wavy line indicates ac (driven) terms. 
    Incoming (resp. outgoing) straight line represent $\hat{a}$ (resp. $\hat{a}^\dagger$) in the represented process, which must conserve energy between drive, incoming and outgoing photons to be resonant. 
    $T$ represents the effective gate duration.}
    \label{tab:diagrams}
	\begin{align*}
		\omega_{\rm r} - \omega_0 
		&= \underset{\begin{tikzpicture} [baseline=(in.north)]
\draw[thick] (0,0)
    node (in) [] {}
    ++ (\ext, 0) node (mid) [vertex] {}
    ++ (0,-\bub) node (bubble) [] {}
    ++ (0, \bub) ++ (\ext, 0) node (out) [] {}
;
\path (in) edge [photon] (mid) ;
\path (bubble.center) edge [out=180, in=-135] (mid) ;
\path (mid) edge [out=-45, in=0] (bubble.center) ;
\path (mid) edge [photon] (out) ;
\end{tikzpicture}}{12g_4^{\rm dc}} 
		- \underset{\begin{tikzpicture} [baseline=(in.north)]
\draw[thick] (0,0)
    node (in) [] {}
    ++ (\ext, 0) node (mid) [vertex] {}
    ++ (0,-\bub) node (up) [vertex] {}
    ++ (0,-\bub) node (bubble) [] {}
    ++ (0, 2*\bub) ++ (\ext, 0) node (out) [] {}
;
\path (in) edge [photon] (mid) ;
\path (mid) edge (up) ;
\path (up) edge [out=-45, in=0] (bubble.center) ;
\path (bubble.center) edge [out=180, in=-135] (up) ;
\path (mid) edge [photon] (out) ;
\end{tikzpicture}\, \begin{tikzpicture} [baseline=(in.north)]
\draw[thick] (0,0)
    node (in) [] {}
    ++ (\ext, 0) node (midL) [vertex] {}
    ++ (\bub, 0) node (midR) [vertex] {}
    ++ (\ext, 0) node (out) [] {}
;
\path (in) edge [photon] (midL) ;
\path (midL) edge [out=45, in=135] (midR) ;
\path (midL) edge [out=-45, in=-135] (midR) ;
\path (midR) edge [photon] (out) ;
\end{tikzpicture}}{60 g_3^{{\rm dc}\, 2}/\omega_{0}}
		+ \dots \\
		K^{(1)}
		&= \underset{\begin{tikzpicture} [baseline=(in.north)]
\draw[thick] (0,0)
    node (inA) [] {}
    ++ (-45: \ext) node (mid) [vertex] {}
    ++ (-135: \ext) node (inB) [] {}
    ++ (45: 2*\ext)  node (outA) [] {}
    ++ (0, -2*\ext/1.41) node (outB) [] {}
;
\path (inA) edge [photon] (mid) ;
\path (inB) edge [photon] (mid) ;
\path (mid) edge [photon] (outA) ;
\path (mid) edge [photon] (outB) ;
\end{tikzpicture}}{12g_4^{\rm dc}}
		- \underset{\begin{tikzpicture} [baseline=(in.north)]
\draw[thick] (0,0)
    node (inA) {}
    ++ (0, -1.41*\ext) node (inB) {}
    ++ (45: \ext) node (midL) [vertex] {}
    ++ (\bub, 0) node (midR) [vertex] {}
    ++ (45: \ext) node (outA) {}
    ++ (0, -1.41*\ext)  node (outB) [] {}
;
\path (inA) edge [photon] (midL) ;
\path (inB) edge [photon] (midL) ;
\path (midL) edge (midR) ;
\path (midR) edge [photon] (outA) ;
\path (midR) edge [photon] (outB) ;
\end{tikzpicture}}{60 g_3^{{\rm dc}\, 2}/\omega_{0}} + \hdots \\
		\frac{\zeta}{\phi_{\rm e}^{\rm ac} T}
		&= \underset{\begin{tikzpicture} [baseline=(in.north)]
\draw[thick] (0,0)
    node (in) {} node {$\scriptstyle\omega_{\rm d}\,\,$}
    ++ (0: \extd) node (mid) [vertex] {}
    ++ (60: \ext) node (outA) {} node {$\scriptstyle\omega_0$}
    (mid) ++ (-60: \ext) node (outB) {} node {$\scriptstyle\omega_0$}
;
\path (in) edge [wave] (mid) ;
\path (mid) edge [photon] (outA) ;
\path (mid) edge [photon] (outB) ;
\end{tikzpicture}}{g_2^{\rm ac}}
		+ \underset{\begin{tikzpicture} [baseline=(in.north)]
\draw[thick] (0,0)
    node (in) {} node {$\scriptstyle\omega_{\rm d}\,\,$}
    ++ (0: \extd) node (mid1) [vertex] {}
    ++ (0: \bub) node (mid2) [vertex] {}
    ++ (60: \ext) node (outA) {} node {$\scriptstyle\omega_0$}
    (mid2) ++ (-60: \ext) node (outB) {} node {$\scriptstyle\omega_0$}
;
\path (in) edge [wave] (mid1) ;
\path (mid1) edge [wave] (mid2) ;
\path (mid2) edge [photon] (outA) ;
\path (mid2) edge [photon] (outB) ;
\end{tikzpicture}}{3g_1^{\rm ac}g_3^{\rm dc}/\omega_{\rm d}} + \dots \\
        \frac{\tau}{\phi_{\rm e}^{\rm ac} T}
		&= \underset{\begin{tikzpicture} [baseline=(in.north)]
\draw[thick] (0,0)
    node (in) {} node [] {$\scriptstyle\omega_{\rm d}\,\,$}
    ++ (0: \extd) node (mid) [vertex] {}
    ++ (60: \ext) node (outA) {} node [] {$\scriptstyle\omega_0$}
    (mid) ++ (0: \ext) node (outB) {} node [] {$\scriptstyle\omega_0$}
    (mid) ++ (-60: \ext) node (outC) {} node [] {$\scriptstyle\omega_0$}
;
\path (in) edge [wave] (mid) ;
\path (mid) edge [photon] (outA) ;
\path (mid) edge [photon] (outB) ;
\path (mid) edge [photon] (outC) ;
\end{tikzpicture}}{g_3^{\rm ac}/2} + \dots
	\end{align*}
\end{table}
In the absence of driving, the effective Hamiltonian contains only number-conserving terms:
\begin{equation}
	\hat{H}_{\rm eff} 
	= (\omega_{\rm r} - \omega_0) \hat{a}^\dagger\hat{a} 
	+ \sum_{n=1}^{\infty} \frac{K^{(n)}}{n+1} \hat{a}^{\dagger n}\hat{a}^n, 
\end{equation}
where $\omega_{\rm r}$ is the renormalized oscillator frequency, $K^{(1)}$ accounts for Kerr effect, and the $K^{(n>1)}$ for higher order effects. Their perturbative expressions are given at Tab.~\ref{tab:diagrams}.
All $K^{(n)}$ are static interactions that cannot be turned off. 
We illustrate their influence on short timescales by considering the evolution of a coherent state $\ket{a}$ centered on $\bar a = \bra{a} \hat{a} \ket{a}$. 
Using a semi-classical approximation, exact for coherent states, and that  $|a|^2$ is a constant of motion, $\bar a$ rotates at the rate 
\begin{equation} \label{eq:drift_angle}
	\mathrm{arg}\,\bar a(t) = -\left(\omega_{\rm r} + \sum_{n=1}^{\infty} K^{(n)} |a|^{2n} \right)t.
\end{equation}
The $|a|^2$ dependence implies that a superposition of coherent states will deform under free evolution.
This clearly undermines our control of the system. 
The main feature of the SNAIL dipole is the possibility to tune the $K^{(n)}$ coefficients \textit{via} $\phi_{\rm e}^{\rm dc}$.
We make use of this control knob to cancel the static nonlinearity, at least in the $|\alpha|^2 \lesssim 1$ region, by imposing $K^{(1)}(\phi_{\rm e}^{\rm dc}) =0$. 

\begin{figure} 
	\centering
    \includegraphics{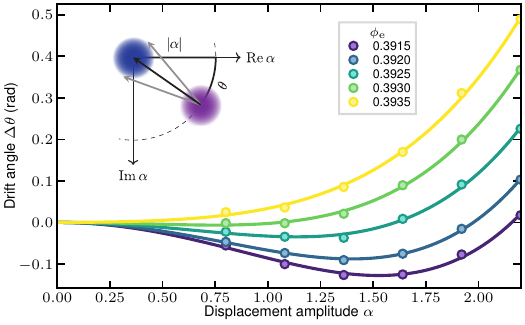}
	\caption{Calibration of the Kerr-free operation point. Drift angle measured as function of $|\alpha|$, for several $\phi_{\rm e}^{\rm dc}$ values after 100 ns free evolution. 
    Solid lines indicate fits according to equation (\ref{eq:drift_angle}), and stars indicate the measured drifted angles. 
    (\textbf{inset}) Sketch of the drift angle measure protocol. The vacuum is displaced by $|\alpha|$ and freely evolves, accumulating a phase $\theta$ (pink). 
    We then search for the displacement that maximizes overlap with the vacuum state (blue), whose angle is $\theta$. }
    \label{fig:angle_drift}
\end{figure}

We experimentally locate this Kerr-free flux point by the out-and-back protocol~\cite{sivak_real-time_2023}, in which we monitor the drift angle $\theta$ of a coherent state of amplitude $a$ under a $t=\SI{100}{\nano\second}$ free evolution. To do so, we select the $\theta$ value that maximizes the vacuum overlap of the state after free evolution and displacement by $-|a|\exp{(-i\theta)}$, as sketched on Extended Data Fig.~\ref{fig:angle_drift} (inset). The free evolution time, here $t=\SI{100}{\nano\second}$, is chosen short enough so that the state stays close to a coherent state.
Both displacement pulses last \SI{10}{\nano\second}.
The result, shown in Extended Data Fig.~\ref{fig:angle_drift},  
demonstrates a \textit{Kerr-free zone} in phase space expanding up to $|a| \simeq 1.5$ for $\phi_{\rm e}/\Phi_0=0.393$. By fitting the drift angle using equation (\ref{eq:drift_angle}), we obtain the experimental value of $K^{(1)}$ plotted in Fig.~\ref{fig:setup}f, which agrees with the perturbative expansion prediction, calculated from $g_n^{\rm dc}$ which values were obtained independently from the fitting procedure described above. 

\subsection{Calibration of effective drive strengths} 
When driving the system, the RWA approach is still valid, but  non number conserving terms appear in the effective Hamiltonian:
\begin{equation}
	-iT\hat{H}_{\rm eff,\,drive}
	= \left( \alpha\hat{a}^\dagger + \frac{\zeta}{2}\hat{a}^{\dagger 2} + \tau\hat{a}^{\dagger 3} + \dots\right) - \mathrm{h.c.}, 
\end{equation}
which correspond to displacement, squeezing, and trisqueezing respectively. 
Their expression in terms of elementary processes are given at Tab.~\ref{tab:diagrams}. 
The effective gate time $T = \int_0^t f(\tau)\mathrm{d}\tau$ where $t$ is the total gate time and $f(\tau)$ is the pulse shape.
Note that the diagrams represent each process' resonance condition: $\omega_{\rm d} = \omega_0$ for displacement, $2\omega_0$ for squeezing, and $3\omega_0$ for trisqueezing.

Since the different drive lines have different amplification, attenuation and potentially unwanted reflectance, it is necessary to calibrate the drive amplitudes $\phi_{\rm e}^{\rm ac}$ independently at each frequency multiple.
We measure $\alpha$, $\zeta$ or $\tau$ via Wigner tomography after flux driving at the corresponding frequency for a range of flux-pulse voltage amplitudes $V^{\rm ac}$. The measurement settings are replicated in simulations and the linear scaling factor $V_0$ to match the corresponding flux amplitude $\phi_{\rm e}^{\rm ac} = V^{\rm ac}/V_0$ is fitted for each measurement of $\alpha$, $\zeta$ or $\tau$, respectively, as shown in Fig.~\ref{fig:interactions}c and \ref{fig:interactions}f.  
Note that the simulation and the experiment still match beyond the linear regime at low $\phi_{\rm e}^{\rm ac}$.
This calibration is especially crucial to match the amplitude of the $1\omega_0$ and $3\omega_0$ in the cubic state protocol below.

\subsection{Experimental cubic phase state calibration} 
Experimental generation of the cubic phase gate requires a precise experimental tune-up to balance the timing, strength, and angles of the drive pulses. All pulses in this work are produced by direct digital synthesis in the Presto platform~\cite{tholen2022a} up to 9 GHz. The $3\omega_0$ pulse at $12.5$ GHz is produced by combining the output from two Presto ports in a physical mixer. Hence, we have full phase control of all pulses. The cubic gate, $\exp(i\gamma ((\hat{a}^\dagger+\hat{a})/\sqrt{2})^3)$, contains terms resonant at $3\omega_0$ as well as cross terms like $\hat{a}^{\dagger 2} \hat{a}$ requiring a drive at $1\omega_0$. To activate the interactions with equal strength, the effective drive amplitudes $\phi_{\rm e}^{\rm ac}$ are calibrated according to the procedure in the previous section.
However, such a $1\omega_0$ flux drive also triggers a displacement via $g_1^{\rm ac}$.
It can be counterbalanced by charge driving, also at $1\omega_0$, with opposite phase and equal amplitude. 
This cancellation must be carefully realized, since $g_1^{\rm ac} / g_3^{\rm ac} \approx 2100$, a small misalignment would generate a significant displacement, well outside the Kerr-free zone. 
The amplitudes of the drives at $1\omega_0$ are first calibrated separately for charge and flux by displacing the vacuum with different amplitudes. We then fit the corresponding Poissonian distributions obtained by probing the qubit with a Fock-0 selective $\pi$-pulses to obtain individual amplitude scalings for the flux and charge drives. 
We then characterize the timing of the two $1\omega$ pulses down to $\sim100ps$, by utilizing the digital delay of the arbitrary waveform generator. 
Finally, we finetune the angle and magnitude of the flux drive such that under both drives, the vacuum stays undisplaced. Angles of the squeezing and trisqueezing drives are calibrated by applying each process separately and fitting the angle of the resulting Wigner states. With all amplitudes, angles, and timings calibrated, we apply the gate sequence of the main text to generate the cubic phase state. 

\subsection{Error analysis}
\begin{figure} 
	\centering
    \includegraphics{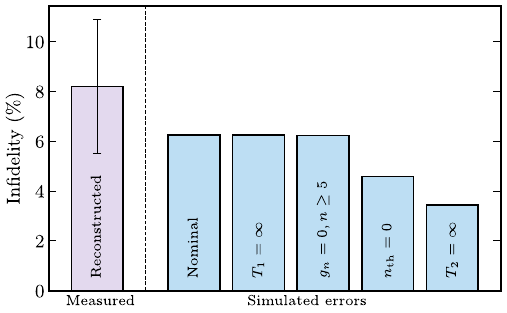}
	
	\caption{Simulated error analysis showing how the infidelity is reduced if an error source is eliminated. Simulated nominal infidelity is obtained by numerical simulation as described in Sec.~\ref{sec:numerical_simulations}, with all parameters at their measured values.}
 \label{fig:error_analysis}
\end{figure}

Measured and similar simulations are made to obtain the infidelity analysis presented in Extended Data Fig.~\ref{fig:error_analysis}, where each bar corresponds to a case with a certain infidelity channels removed. 
The infidelity in each case is defined as the infidelity with respect to the ideal cubic state whose $\zeta$ and $\gamma$ parameters maximize fidelity to the simulated state.  The error budget is found to be non-additive and nonlinear. For example, setting $g_n=0,\ n\geq5$ does reduce the infidelity if both thermal and dephasing are eliminated first. Note that $g_3^{\rm dc}$ and $g_4^{\rm dc}$ also generate contributions on higher orders. The error bar of the measured infidelity is obtained by dividing the data into 11 chronological batches (each $> 1$ hour of averaging), reconstructing each batch and calculating the fidelity with respect to the best fit to all averaged batches. Part of the measured infidelity is found to originate from drift of the system which is observed as a slightly tilting cubic phase state when the system slightly deviates from the Kerr-free point. Note that the slow drift of the dc-flux offset is not considered in the simulation, but is an important factor in the difference between the collected data batches.

\subsection{Dephasing noise} 
\begin{figure}
    \centering
    \includegraphics{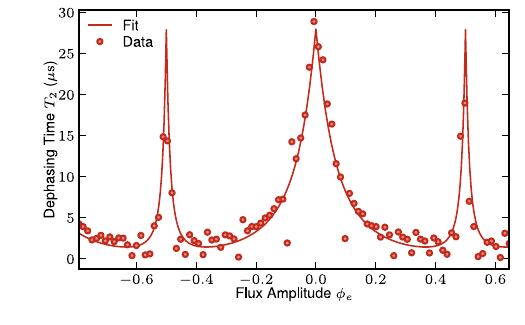}
    \caption{The cavity $T_2$ measured by Ramsay interference protocol as a function of external flux. The data are fitted by a model including $T_1$, broadband and $\nicefrac{1}{f}$ flux noise.}
    \label{fig:supp:T2}
\end{figure}

The main source of decoherence for our system is flux noise, which induces fluctuations in the resonator frequency, and thus dephasing.
According to this noise mechanism, the $T_2$ time should strongly depends on the $\omega_0$ dependence \textit{w.r.t.} $\phi_{\rm e}$. 
Following \cite{mergenthaler_2021_EffectsSurface,yoshihara_2006_DecoherenceFlux}, we assume the flux noise power spectrum to be composed of a $1/f$ part and a broadband part, $S_{\phi_{\rm e}}(\omega) = 2\pi A_{\nicefrac{1}{f}}/\omega + S_{\rm bb}$, which lead to
\begin{equation}
	\frac{1}{T_2} 
	= \frac{1}{2T_1} 
	+ 2\pi \sqrt{2\log(2) A_{\nicefrac{1}{f}}} \frac{\mathrm{d} \omega_0}{\mathrm{d}\phi_{\rm e}}
	+ 2\pi S_{\rm bb} \left(\frac{\mathrm{d}\omega_0}{\mathrm{d}\phi_{\rm e}}\right)^2.
\end{equation}
The fit to experimental data in Extended Data~\figref{fig:supp:T2} shows good agreement with this model. 
One notes the existence of sweet spots protected from dephasing at $\phi_{\rm e}=0$ and $1/2$. 
Since they do not coincide with the Kerr free flux point found at $\phi_{\rm e} \simeq 0.3930$, they cannot be exploited in our current device. 
Optimizing the sweet spot position to enhance $T_2$ is a promising avenue to enhance the fidelity of any operation.

\subsection{Device fabrication}
The device is fabricated from e-beam evaporated Aluminium on a \SI{330}{\micro\meter} thick C-axis sapphire. The large features are patterned with an optical lithography and followed by a wet etch. The junctions are added with e-beam lithography and attached to the base layer with a separate patch layer~\cite{burnett2019}. The SNAIL has three junctions in series in one arm, and one single junction in the other arm. The junctions are fabricated with a variation of the Manhattan technique that allows for arbitrarily large junctions~\cite{Raftery2017}. Based on a room temperature resistance measurement of similar junctions, the critical current of a large junction is \SI{434}{\nano\ampere}, while it is \SI{46}{\nano\ampere} for the small junction giving an asymmetry factor $\beta = 0.106$ , close to the fitted value of 0.097.

\subsection{Author contributions}
A.M.E. and M.Ke. built the experimental setup.
A.M.E. carried out the experiments.
M.Ke. and C.C.M. fabricated the chip.
T.S. and T.H. developed the theory. 
A.M.E. and T.S. analyzed the data.
A.M.E. and S.D. developed the measurement software framework.
Y.L. and J.Y optimized and characterized an earlier version of the chip.
T.S., M. Ke, T.H. and M. Ka performed numerical simulations.
A.M.E., T.S., T.H., M.Ke., M.Ku., M.Ka., P.D., and S.G. regularly discussed the project and provided insight.
S.G. supervised the project.
A.M.E., T.S., T.H., M.Ke., and S.G. wrote the manuscript with feedback from all authors.

\begin{acknowledgments}
A.M.E would like to thank Giulia Ferrini for valuable discussions.
The simulations and visualization of the quantum states were performed using QuTiP. The chips were fabricated in the Chalmers Myfab cleanroom. This work was supported by the Knut and Alice Wallenberg Foundation via the Wallenberg Centre for Quantum Technology (WACQT) and by the Swedish Research Council. T.H. also acknowledges the financial support from the Chalmers Excellence Initiative Nano. 

\end{acknowledgments}

\bibliography{refs}
\onecolumngrid
\clearpage

\end{document}


\title{Universal control of a bosonic mode via drive-activated native cubic interactions --  Supplementary Material}
\maketitle
\setcounter{table}{0}
\renewcommand{\thetable}{S\arabic{table}}
\setcounter{figure}{0}
\renewcommand{\thefigure}{S\arabic{figure}}

\setcounter{section}{0}
\renewcommand{\thesection}{S\arabic{section}}
\setcounter{equation}{0}
\renewcommand{\theequation}{S\arabic{equation}}

\setcounter{page}{1}

\subsection{Hardware design and setup}

\begin{figure}[!t]
\includegraphics[width=\textwidth]{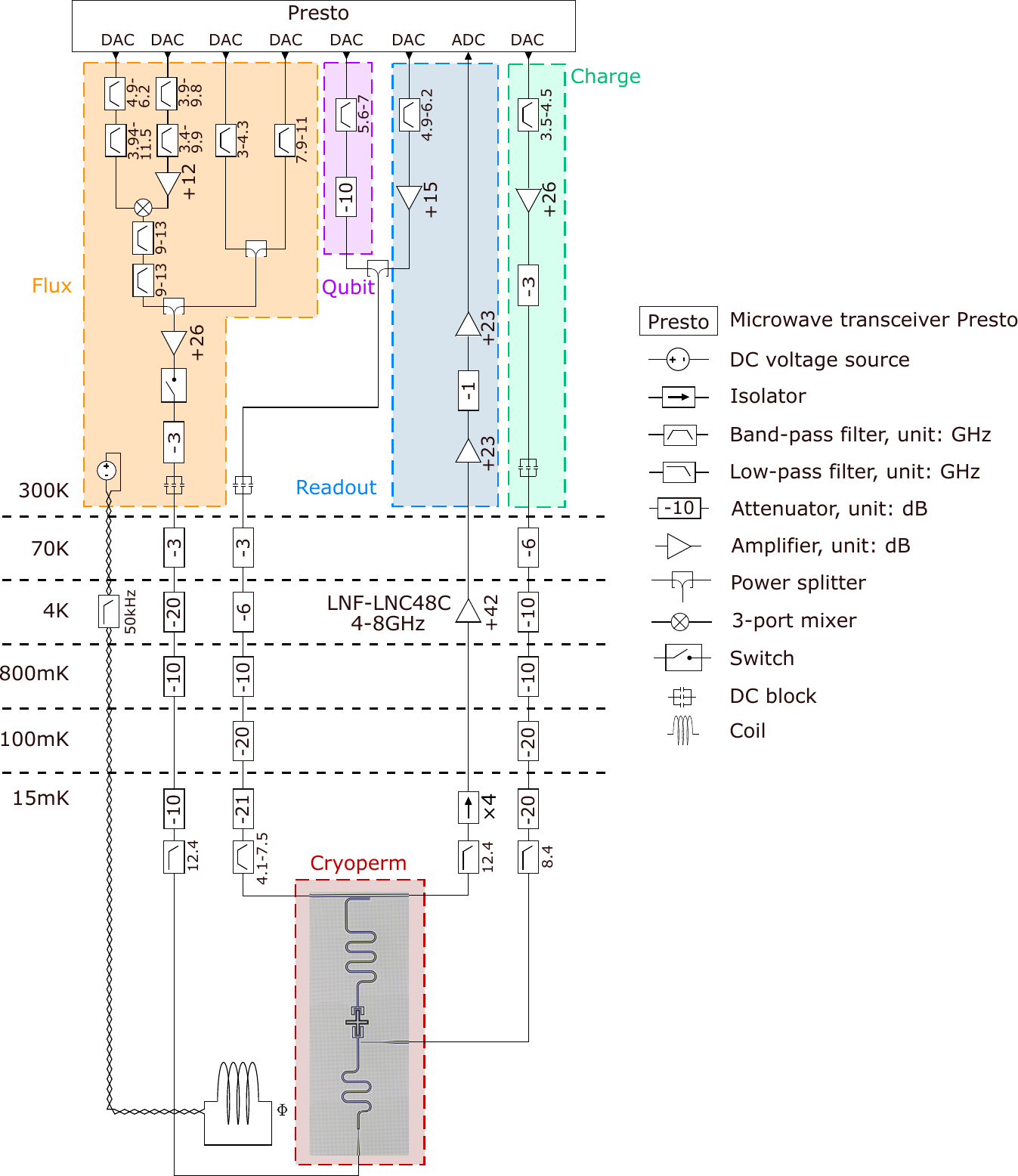}
\caption{\label{figSI:SI_setup} Hardware setup of the experiment.
}
\end{figure}
To control and measure the device, we use the microwave transceiver Presto~\cite{tholen2022a}. By utilizing the second Nyquist band, Presto performs direct digital synthesis up to $\sim 9$ GHz with an instantaneous bandwidth of $\pm$500 MHz. The outputs are then filtered with external band-pass filters, which cut the unwanted signals at the other Nyquist bands. The fundamental frequency of the SNAIL-resonator was chosen such that both the $1\omega_0=4.1582$ and the  $2\omega_0=8.3164$ drive pulses can be generated directly. To synthesize the $3\omega_0$ pulses, we generate one tone from Presto as a constant local oscillator at 7.9 GHz and mix it with another pulse-shaped tone in a three-port mixer ZX05-153-S+. Thereby, we have full phase control of all pulses used in the experiment. The input lines are first amplified at room temperature, and then attenuated at the different stages of the cryostat, and finally filtered at the mixing chamber. The output signal is amplified by a HEMT amplifier at 4 K, and two low-noise amplifiers at room temperature before being digitized by Presto. The full wiring diagram is illustrated in Fig.~\ref{figSI:SI_setup}.

The system parameters are displayed in Tab.~\ref{tab:system_param}. The thermal population gets considerably enhanced by noise from the flux line amplifiers. To mitigate this, we use a microwave switch ADRF5024 which cuts the line when flux pulsed are not sent. With the switch, we are able to suppress the SNAIL-resonator thermal population to $2.4 \%$. The population is characterized by the ratio between a SNAIL-resonator Fock level selective pulse on Fock 1 compared to Fock 0. The thermal population could potentially be decreased further by optimizing the duration during which the switch is closed.

The dispersive shift between the SNAIL-resonator and the qubit is \SI{2.34}{\mega \hertz} which allows for SNAIL-resonator Fock-level selective pulses on the qubit as well as Wigner tomography, which utilizes Fock level inselective pulses. 

\begin{table}[]
\caption{\label{tab:system_param}
System parameters}
\begin{ruledtabular}
\begin{tabular}{llr}
Parameter & Symbol & Value \\ 
\hline
SNAIL-resonator frequency & $\omega_{\rm r}/2\pi$ & \SI{4.1582}{\giga\hertz} \\ 
SNAIL-resonator relaxation time & $T_{1c}$ & \SI{28(3)}{\micro\second} \\
SNAIL-resonator coherence time & $T_{2c}^*$ & \SI{2.80(8)}{\micro\second} \\
SNAIL-resonator thermal pop. & $P_c$ & \num{2.4}\% \\
SNAIL-resonator impedance & $Z$ & \SI{57.94}{\ohm} \\ 
SNAIL-resonator-qubit disp.~shift & $\chi_{cq}/2\pi$& \SI{2.34}{\mega\hertz}\\
SNAIL-asymmetry & $\beta$ & \num{0.097} \\
SNAIL large junction Josephson energy & $E_{\rm J}$ & \SI{245}{\giga\hertz} \\
Free resonator frequency & $\omega_{\infty}/2\pi$ & \SI{8.99}{\giga\hertz} \\

Qubit frequency & $\omega_q/2\pi$ & \SI{5.2891}{\giga\hertz} \\
Qubit relaxation time & $T_{1q}$ & \SI{14(5)}{\micro\second} \\
Qubit decoherence time & $T_{2q}^*$ & \SI{9.1(3)}{\micro\second} \\
Qubit anharmonicity & $\alpha_{q}/2\pi$& \SI{236}{\mega\hertz} \\
Qubit thermal pop. & $P_q$ & \num{1.3}\% \\

Readout frequency & $\omega_r/2\pi$ & \SI{6.8087}{\giga\hertz} \\
\label{table:system}
\end{tabular}
\end{ruledtabular}
\end{table}

\subsection{Harmonic analysis}
\begin{figure}[b!]
	\centering
	\includegraphics{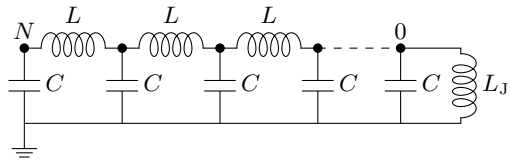}
	\caption{Discretized, linearized model of the SNAIL-resonator. }
	\label{fig:telegraph_chain}
\end{figure}
In this section, we explicitly derive the $\omega_0$ and $\Phi$ parameters, respectively the SNAIL-resonator bare frequency and the participation of the SNAIL flux to the resonance respectively. 
They are obtained by harmonic analysis of the SNAIL-terminated resonator system. 
To do so, we first neglect all $n>2$ non-linear terms in the SNAIL potential Taylor series, which amounts to replacing the SNAIL by a linear inductor of inductance
\begin{equation} \label{eq:inductance}
	1/L_{\rm J} = E_{\rm J}\left(\beta\cos\varphi_{\rm m}
	+ \frac{1}{n} \cos\left(\frac{\phi_{\rm e} - \varphi_{\rm m}}{n}\right)\right)
\end{equation}
at the current anti-node of the $\lambda/4$ resonator. 
To treat the boundary condition, one can discretize the resonator as a series of elementary $LC$-oscillators, with phases labeled from $\varphi_0$ at the SNAIL boundary to $\varphi_{N}$ at the open edge, as shown on \figref{fig:telegraph_chain}. 
the corresponding Lagrangian is~\cite{vool_2017_IntroductionQuantum, rasmussen_2021_SuperconductingCircuit}
\begin{equation}\label{eq:discrete_lagrangian}
	\mathcal{L} =
	\frac{1}{2}\sum_{i,j=0}^{N} C \dot\varphi_i^2
	- \frac{1}{L}\varphi_i M_{ij}\varphi_{j}, 
    \qquad M = \begin{bmatrix}
		1+\frac{L}{L_{\rm J}} & -1              \\
		-1                    & 2  & -1         \\
		                      &    & \ddots     \\
		                      &    &        & 1
	\end{bmatrix}.
\end{equation}
We diagonalize the model by $M_{ij} = \sum_k P_{ik} \lambda^2_k P_{kj}$, and change the variables as $\sum_j P_{kj}\varphi_j = \phi_k$, to reach a Lagrangian of non-interacting modes.
It is Legendre transformed to reach the Hamiltonian
\begin{equation}
	\mathcal{H} = \sum_{k=0}^N \frac{1}{2C} n_k^2 + \frac{\lambda_k^2}{L}\phi_k^2.
	\qquad (n_k=C \dot\phi_k)
\end{equation}
We perform canonical quantization and introduce creation/annihilation operators as $\hat a_k = (\hat \phi_k + i Z_k \hat n_k)/\sqrt{2Z_k}$, where $Z_k = \sqrt{L/C}/\lambda_k$ and $\omega_k = \lambda_k/\sqrt{LC}$.
The resulting Hamiltonian is simply
\begin{equation}
	\hat H
	= \sum_{k=0}^N \omega_k \hat{a}_k^\dag\hat{a}^{\phantom{\dag}}_k
	+ U_{\rm NL}(\hat\varphi_0),
	\qquad \hat\varphi_0
	= \sum_{k=0}^N P_{k0} \sqrt{\frac{Z_k}{2}} (\hat a^\dag_k + \hat a_k),
\end{equation}
where we reintroduced the non-linear part of the SNAIL potential $U_{\rm NL}(\hat\varphi_0)$.
We now focus on the fundamental mode by truncating all sums to $k=0$.
Our goal is to express $\lambda_0$ and $P_{00}$ in terms of $\phi_{\rm e}$ and other circuit parameters.
This is obtained by explicit diagonalization of $M$ (\textit{cf.}~\eqref{eq:discrete_lagrangian}) in the continuum limit, which consists of $N \to \infty$ and $L, C \to 0$, while $v = d/(N \sqrt{LC})$ the wave velocity, and $Z=\sqrt{L/C}$ the resonator impedance, are kept constant.

The diagonalization is helped by introducing a plane wave ansatz, $P_{lj} = N_k \cos(j k_l + \delta_l)$.
Injecting it in the eigenvalue equation, we obtain three conditions,
\begin{equation}
	\lambda_l \simeq k_l,
	\quad \tan \delta_l = \frac{L}{L_{\rm J}}\frac{1}{k_l},
	\quad \tan(Nk_l + \delta_l) = 0.
\end{equation}
They correspond respectively to the dispersion relation and left and right boundary conditions.
We deduce the implicit equation for the fundamental frequency $\omega_0$,
\begin{equation} \label{eq:resonance_frequency}
	\frac{\omega_0}{\omega_\infty}
	= \frac{2}{\pi}\arctan\left(\frac{Z}{L_{\rm J} \omega_0}\right),
	\quad \omega_\infty = \frac{\pi}{2}\frac{v}{d},
\end{equation}
where $\omega_\infty$ is the mode frequency in the absence of the SNAIL, $1/L_{\rm J} \to \infty$. This equation is solved numerically to obtain $\omega_0$. Note that $Z$ and $E_{\rm J}$ do not appear independently in this equation. 
They rather form a single free parameter $Z/E_{\rm J}$.

The last unknown is the normalisation $N_0$. 
It is obtained from the orthogonality condition $\sum_i P_{ki} P_{ki} = 1$, which provides in the continuum limit
\begin{equation}
	N_0^{-2} = N \int_0^1 \mathrm{d}x\,
	\cos^2\left(x \frac{\pi}{2}\frac{\omega_0}{\omega_\infty} + \delta_0\right),
\end{equation}
which is integrated into
\begin{equation}
	\hat \varphi_0
	= \sqrt{\frac{Z}{\pi \omega_0/\omega_\infty} \frac{1+\cos(\pi \omega_0/\omega_\infty)}{1+\mathrm{sinc}(\pi\omega_0/\omega_\infty)}}
	(\hat a^\dag_0 + \hat a_0).
\end{equation}
We have reached $\Phi$ defined in Methods section A. Note that in the main text, the mode index is dropped.

\subsection{Pulse calibration}
\begin{figure}
    \centering
    \includegraphics[scale=1]{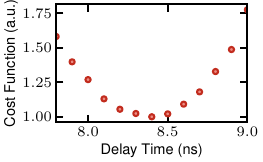}
    \caption{Time-delay calibration of the $1\omega$ flux and charge drives.  }
    \label{fig:delay}
\end{figure}

The $1\omega_r$ flux and charge pulses need to be carefully calibrated to capture the third order cross terms while avoiding the linear displacement. To calibrate the pulses, we apply them simultaneously in (close to) opposite directions with (close to) equal strengths. If the pulses are ill timed, the vacuum will displace (by the pulse arriving first) to large amplitudes where it gets distorted by higher order Kerr corrections before the other pulse displaces the state back to the origin. The cost function is the normalized mean squared distance of the measured Wigner pixels with respect to the vacuum state, which is optimal when the two pulses are timed at $8.4$ns (Fig.~\ref{fig:delay}).

\bibliography{refs}